\documentclass[twocolumn,showpacs,amsmath,amssymb,floatfix,pre]{revtex4}
\usepackage{graphicx}
\begin{document}
\title{Solution of a model of self-avoiding walks with multiple
  monomers per site on the Bethe lattice} 
\author{Pablo Serra}
\email{serra@famaf.unc.edu.ar}
\affiliation{Facultad de Matem\'atica, Astronom\'{\i}a y F\'{\i}sica\\
Universidad Nacional de C\'ordoba\\
C\'ordoba - RA5000\\
Argentina}
\author{J\"urgen F. Stilck}
\email{jstilck@if.uff.br}
\affiliation{Instituto de F\'{\i}sica\\
Universidade Federal Fluminense\\
Av. Litor\^anea s/n\\
24210-346 - Niter\'oi, RJ\\
Brazil}
\date{\today}

\begin{abstract}
We solve a model of self-avoiding walks with up to two monomers per
site on the Bethe lattice. This model, inspired in the Domb-Joyce model,
was recently proposed to describe the collapse transition observed in
interacting polymers \cite{kpor06}. When immediate self-reversals are
allowed (RA model), the solution
displays a phase diagram with a
polymerized phase and a non-polymerized phase, separated by a phase
transition which is of first order for a non-vanishing statistical weight
of doubly occupied sites. If the configurations are restricted
forbidding immediate self-reversals (RF model), a richer phase diagram with
two distinct polymerized phases is found, displaying a tricritical
point and a critical endpoint.
\end{abstract}

\pacs{05.40.Fb,05.70.Fh,61.41.+e}

\maketitle

\section{Introduction}
\label{intro}
Since the pioneering work by Flory \cite{f66} self- and mutually
avoiding walks on lattices have been used extensively to model linear
polymer molecules. The realization that the graphs relevant in the
high-temperature series expansion of the n-vector model of magnetism
correspond to self-avoiding walks in the limit $n \to 0$ \cite{dg72}
made it possible to apply the ideas of the renormalization group to
polymer systems, and the concept of scaling has been very important in
the later development of this field \cite{dg79}.

If the polymer chains are placed in a solvent, as the temperature is
lowered the chains may change from an extended to a collapsed
configuration. The temperature where this transition happens was named
the $\theta$ temperature \cite{f66}. This phenomenon was modelled by
self-avoiding walks with an attractive interaction between monomers on
first neighbor sites of the lattice (interacting SAW's - ISAW), and
if this problem is considered in an 
ensemble which is grand-canonical with respect to the number of
monomers the $\theta$-point of polymer
collapse was found to be a tricritical point \cite{dg75}. Thus, in these
generalized parameter space, the transition between a non-polymerized
and a polymerized phase is continuous at high temperatures, but
becomes of first order as the temperature is lowered, and a
tricritical point separates these two transition lines. The
polymer-solvent system may also be treated in more detail, using a
lattice gas model with two different particles (polymer monomers and
solvent molecules) \cite{wp81}, and such models describe quite well
experimental results of equilibrium polymerization in solution
\cite{ks84}. Actually the simpler ISAW model may be obtained as a
limiting case of the more general diluted polymer model
\cite{mz90,ss90}, and the tricritical points in the two models
apparently belong to the same universality class. In two dimensions,
extensive numerical work has been done to calculate the tricritical
exponents of the ISAW model \cite{ds85,s86,mz90}, and for the diluted
polymer model the tricritical exponents were calculated exactly
\cite{ds87,d88}. Also, it was found that in some cases the ISAW model
may display a third phase, besides the non-polymerized and the
polymerized phases, which is a dense phase. This behavior seems to be
restricted to two-dimensional lattices, and was found on the Husimi
lattice \cite{sms96,p02,ssm02,sscm04} as well as on the square
lattice, using transfer matrix and finite size scaling techniques
\cite{mos01}. From these studies one may conclude that even
qualitative aspects of the thermodynamic behavior of interacting
polymers may be determined by details of the model. On the square and
the $q=4$ Husimi lattice, for example, a dense phase exists if the
interaction is assumed to be between first-neighbor bonds, but is not
stable if interacting monomers are considered.

Including interactions between SAW's is a major complication of the
model, both in closed form approximations and in transfer matrix
calculations. In a recent paper \cite{kpor06}, Krawczyk
et. al. proposed and alternative approach which seems to be very
promising, since only one-site interactions are present. In this
model, $n_i=0,1,\ldots,K$ monomers may be placed on site $i$ of the
lattice, and each site contributes with a factor $\omega_{n_i}(i)$ to
the statistical weight of a configuration. Without loss of
generalization, we may fix the weight of an empty site to be equal to
one. This model is inspired on a
generalization of a model of weighted random walks proposed by Domb
and Joyce some time 
ago \cite{dj72} and a it was also discussed in \cite{dq85} for the
particular choice $\omega_{n_i}=\omega^{-n_i}$ of the weight
functions. The model was studied in \cite{kpor06} using 
simulations for the particular case $K=3$ and as a function of
$\omega_2$ and $\omega_3$, fixing $\omega_1=1$. With a fixed value for
$\omega_1$ only polymerized phases will appear, and the authors look
for transitions between those phases. For the particular case of a
model where immediate self-reversals are forbidden, simulations on a
cubic lattice (model RF3) show a transition between extended and
collapsed phases. As the parameters are changed, there are indications
in the simulations that the transition changes from second to first
order. In another case (RA2-immediate reversals allowed, simulations
on the square lattice), no evidence of a phase transition is found.

In this paper, we solve the model proposed in \cite{kpor06} on a Bethe
lattice with 
coordination number $q$. Although the critical exponents are classical
for Bethe lattice solutions, they may provide better information on
the thermodynamic behavior of models than simpler mean-field
approximations, particularly for polymer models, since first neighbor
correlations are included \cite{sw87}. For the sake of simplicity, we
restrict our calculation to the case $K=2$, where collapse may already
occur, but we will not assume $\omega_1=1$, so that transitions between
polymerized and non-polymerized phases may be found. Similar to what
happens in the studies of the ISAW model and in the simulations of
the model with multiple monomers per site \cite{kpor06}, qualitative
differences are found on the Bethe lattice solutions of the RA and RF
models. In the more restrictive RF model, we find two distinct
polymerized phases, one of them with double occupied sites only. This
phase does not appear in the RA model. For $q>2$, no qualitative
differences in the phase diagrams are found as $q$ is
changed. Although a tricritical point is found in the RF model, as
expected, in the RA model the transition between the non-polymerized
and the polymerized phases is always discontinuous for non vanishing
values of $\omega_2$. 

In section \ref{defmod} the model is defined in greater detail and its
solution on the Bethe lattice in terms of recursion relations is
presented. The thermodynamic properties of the model are obtained from
the recursion relations and presented in section \ref{tpm}. Final
conclusions and discussions are in section \ref{con}.

\section{Definition of the model and solution in terms of recursion
  relations}
\label{defmod}
We consider linear polymers, formed by monomers linked by bonds. The
monomers are placed on the sites of a lattice and the chains are self-
and mutually avoiding walks on this lattice. Up to $K$ monomers may be
located at the 
same lattice site. All monomer are supposed to be distinguishable. The
polymeric chains are linear self- and mutually-avoiding walks, each
step linking two monomers located on first-neighbor sites. Thus
multiple steps of the walk may be placed on the same bond of the
lattice.  The chains are not allowed to form rings. The statistical
weight of a configuration will be a product 
over site configuration weights $\omega_{n_i}(i)$, where
$i$ labels the site and $n_i=0,1,\ldots,K$ is the number of monomers
located on site $i$. We assume the weight of an empty site to be
unitary ($\omega_0(i)=1$) and the weights to be independent of the
site ($\omega_{n_i}(i)=\omega_{n_i}$). The model may be interpreted as an
interacting polymer model, so that $\omega_1=z=\exp(\mu/k_BT)$ is the
fugacity of a monomer and $\omega_n=z^n\,\exp(-\epsilon(n)/k_BT)$,
$\epsilon_n$ being the interaction energy of the set of $n$ monomers
located on the same site. Monomers on different sites do not
interact. Usually, if we are interested in the collapse transition, the
monomer-monomer interactions should be attractive, and thus
$\epsilon(n)<0$. For $K=1$ the usual polymerization model in the
grand-canonical ensemble is recovered, which has been much studied as
a model for equilibrium polymerization \cite{wp81}. The model
described here is the same studied by 
Krawczyk et al \cite{kpor06}, with a slight generalization, since they
fixed $z=1$ in their simulations. With the parametrization
adopted by them, the polymerization transition between a
non-polymerized and a polymerized phase is not found. Following
Krawczyk et al \cite{kpor06}, we consider two versions of the
model. In the RA (reversion allowed) variant, all walks which obey the
self-avoidance restraint are included, whereas in the RF (reversal
forbidden) variant, only walks without immediate self-reversals are
allowed. In other words, in the RF case 
configurations such that the walk reaches a site and immediately
leaves this site through the same bond are prohibited.

\begin{figure}
\begin{center}
\includegraphics[height=6.0cm]{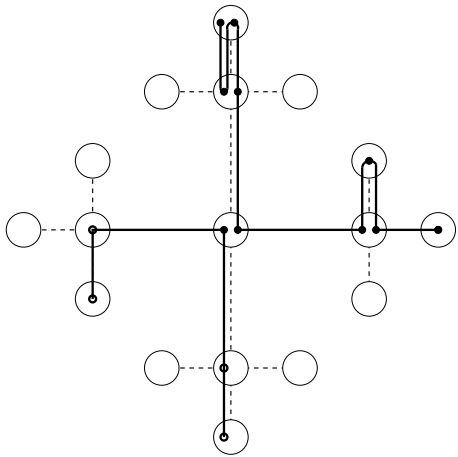}
\caption{Example of a configuration of the Cayley tree with $q=4$ and
  3 generations. This configuration is allowed in the RA model, but
  not in the RF model. Polymer bonds (steps of the walks) are
  represented by full lines, while the lattice bonds are dashed
  lines.} 
\label{f1}
\end{center}
\end{figure}

We will here discuss the 
solution of the model for the particular case
$K=2$ on the Bethe lattice. This is the smallest value for $K$ for
which the collapse may be found, so we have chosen it for
simplicity. The Bethe lattice is the core of a Cayley tree,
and the exact solution of statistical models on it may be viewed as an
approximation of the solutions of the same models on regular lattices
with the same coordination number \cite{b82}. We consider a Cayley
tree of coordination number $q$ and place the endpoints of the walks
on the surface of the tree. In Fig. \ref{f1} a configuration  which
contributes to the partition function of the RA model is
shown for a tree with $q=4$ and three generations. This walk would not
be considered in the RF model, since two immediate self-reversals are 
present. The statistical weight of the depicted configuration is
$\omega_1^6\omega_2^4$.

\begin{figure}
\begin{center}
\includegraphics[width=6.0cm]{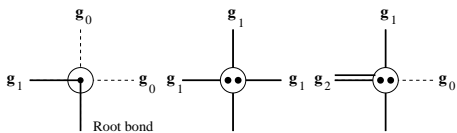}
\caption{Contributions to the recursion relation for the partial
  partition function $g_1$ (RF model).}
\label{f2}
\end{center}
\end{figure}

The solution of the model on the Bethe lattice is obtained by defining
partial partition functions on rooted sub-trees, and realizing that a
subtree with $n+1$ generations may be obtained through the connection
of $q-1$ $n$-generations subtrees to a new site and bond
\cite{sw87}. This operation leads to recursion relations for the
partial partition functions. We start with the more restrictive RF
model. One 
should be careful with the multiplicity of each contribution to the
recursion relations, recalling that even monomers placed on the same
site of the lattice should be considered distinguishable. The
partial partition functions for the model are $g_0$, $g_1$, and $g_2$,
where the subscript denotes the number of polymer bonds on the root
bond of the subtree. In Fig. \ref{f2} the contributions to the recursion
relation for $g_1$ are shown graphically, in the order they appear
in the equations \ref{gp1}. 

Below the recursion relations are presented. In general, we have
$g'_i=\sum_j g'_{i,j}$, and, whenever appropriate, the terms in the
sums begin with a product of two numerical factors, the first of which
is the multiplicity of the configuration of the incoming bonds and the
second is the multiplicity of the connections with the monomers
located at the new site. The recursion relation for $g'_0$ is:
\begin{subequations}
\begin{eqnarray}
g'_{0,1} &=&   g_0^{q-1}, \\
\mbox{}  \nonumber  \\
g'_{0,2}\,&=&\, \binom{q-1}{2} \times 1 \;  \omega_1 \,g_0^{q-3} \,g_1^2,
\\
\mbox{}  \nonumber  \\
g'_{0,3}\,&=&\, \binom{q-1}{4} \times 6 \; \omega_2 \,g_0^{q-5} \,g_1^4 ,\\
\mbox{}  \nonumber  \\
g'_{0,4}\,&=&\, 3\,\binom{q-1}{3} \times 4   \; \omega_2 \,g_0^{q-4}
\,g_1^2 \,g_2,\\ 
\mbox{}  \nonumber  \\
g'_{0,5}\,&=&\, \binom{q-1}{2} \times 4 \; \omega_2 \,g_0^{q-3} \,g_2^2.
\end{eqnarray}
\end{subequations}
The terms of the recursion relation for $g'_1$ are:
\begin{subequations}
\begin{eqnarray}
g'_{1,1}\,&=&\, (q-1) \times 1  \; \omega_1 \,g_0^{q-2} \,g_1, \\
\mbox{}  \nonumber  \\
g'_{1,2}\,&=& \binom{q-1}{3} \times 6 \; \omega_2 \, g_0^{q-4}
\,g_1^3 , \\
\mbox{}  \nonumber  \\
g'_{1,3}\,&=&\, 2 \binom{q-1}{2} \times 4 \; \omega_2 \,g_0^{q-3}
\,g_1 \,g_2.
\end{eqnarray}
\label{gp1}
\end{subequations}
For $g'_2$ we find two terms:
\begin{subequations}
\begin{eqnarray}
g'_{2,1}\, &=& \binom{q-1}{2} \times 2 \; \omega_2 \,g_0^{q-3} \,g_1^2,\\
\mbox{}  \nonumber  \\
g'_{2,2}\,&=&\, (q-1) \times 2 \; \omega_2  \,g_0^{q-2} \,g_2.
\end{eqnarray}
\label{rrg2rf}
\end{subequations}

We may now define ratios of the partial partition functions $R_1=g_1/g_0$ and 
$R_2=g_2/g_0$. Any expected value in the center of the tree may be
expressed as a function of these ratios, so that if we are interested
in the behavior of the model in the center of the tree and in the
thermodynamic limit, we should find the fixed point of the recursion
relations for the ratios, which are:
\begin{subequations}
\begin{eqnarray}
R'_1 \,&=&\, \left[  (q-1) \, \omega_1 \,+\,
 \, 6 \binom{q-1}{3} \,\omega_2 \, R_1^2 \,+\,
\right. \nonumber \\ 
\mbox{}  \nonumber  \\
&& \left. 8 \binom{q-1}{2}\,  \omega_2 \,R_2\, \right]
\,\frac{R_1}{D},   \\ [4ex]
%\mbox{}  \nonumber\\
R'_2 \,&=& \,2 \,\left[\binom{q-1}{2} \,R_1^2 \,+\,
 (q-1) \, R_2 \,\right]\,\frac{\omega_2}{D},
\end{eqnarray}
\label{rrrf}
\end{subequations}
where
\begin{eqnarray}
D\,&=&\,1\,+\, \binom{q-1}{2} \omega_1  \,R_1^2 \,+\,
 \, 6 \binom{q-1}{4}\,\omega_2 \,R_1^4 \, +\nonumber \\
\mbox{}  \nonumber  \\
&& 12 \binom{q-1}{3}\,\omega_2  \,R_1^2 \,R_2\,+\,
 4 \binom{q-1}{2}\,\omega_2  \,R_2^2.
\end{eqnarray}

For the RA model, where immediate self-reversals are allowed, three
other root configurations for subtrees are possible. The
additional partial partition functions are: $g_3$ (two polymer bonds at
the root, connected to the same monomer above), $g_4$ (same as for
$g_3$, but with more than one monomer in the closed path), and $g_5$
(three polymer bonds at the root, two of them connected to the same
monomer above). The recursion relations for this generalized model are:
\begin{subequations}
\begin{eqnarray}
g'_{0,1}\,&=&\, g_0^{q-1},\\
\mbox{}  \nonumber  \\
g'_{0,2}\,&=&\, \binom{q-1}{2} \times 1 \; \omega_1 \,g_0^{q-3}
\,g_1^2,\\ 
\mbox{}  \nonumber  \\
g'_{0,3}\,&=&\, \binom{q-1}{4} \times 6 \; \omega_2 \,g_0^{q-5} \,g_1^4 ,\\
\mbox{}  \nonumber  \\
g'_{0,4}\,&=&\, 3\,\binom{q-1}{3} \times 6   \; \omega_2 \,g_0^{q-4}
\,g_1^2 \,g_2,\\ 
\mbox{}  \nonumber  \\
g'_{0,5}\,&=&\, \binom{q-1}{2} \times 6 \; \omega_2 \,g_0^{q-3}
\,g_2^2,\\ 
\mbox{}  \nonumber  \\
g'_{0,6}\,&=&\, (q-1) \times 1 \; \omega_1 \, g_0^{q-2}g_2,\\
\mbox{}  \nonumber  \\
g'_{0,7}\,&=&\, 3\,\binom{q-1}{3} \times 2 \; \omega_2\,g_0^{q-4} \,
g_1^2 \, g_3, \\
\mbox{}  \nonumber  \\
g'_{0,8}\,&=&\, 3\,\binom{q-1}{3} \times 4 \; \omega_2\,g_0^{q-4} \,
g_1^2 \, g_4, \\
g'_{0,9}\,&=&\, 2\,\binom{q-1}{2} \times 2 \; \omega_2\,g_0^{q-3} \,
g_2 \, g_3, \\
\mbox{}  \nonumber  \\
g'_{0,10}\,&=&\, 2\,\binom{q-1}{2} \times 4 \; \omega_2\,g_0^{q-3} \,
g_2 \, g_4, \\
\mbox{}  \nonumber  \\
g'_{0,11}\,&=&\, 2\,\binom{q-1}{2} \times 2 \; \omega_2\,g_0^{q-3} \,
g_1 \, g_5.
\end{eqnarray}
\end{subequations}
Notice that, although the first five terms are similar to the ones in
the RF model, the multiplicities related to the connections of the two
additional monomers may be larger in the less restrictive RA
model. The contributions to $g'_1$ are:
\begin{subequations}
\begin{eqnarray}
g'_{1,1}\,&=&\, (q-1) \times 1  \; \omega_1 \,g_0^{q-2} \,g_1, \\
\mbox{}  \nonumber  \\
g'_{1,2}\,&=& \binom{q-1}{3} \times 6 \; \omega_2 \, g_0^{q-4}
\,g_1^3 , \\
\mbox{}  \nonumber  \\
g'_{1,3}\,&=&\, 2 \binom{q-1}{2} \times 6 \; \omega_2 \,g_0^{q-3}
\,g_1 \,g_2,\\
\mbox{}  \nonumber  \\
g'_{1,4}\,&=&\, 2 \binom{q-1}{2} \times 2 \; \omega_2 \,g_0^{q-3}
\,g_1 \,g_3,\\
\mbox{}  \nonumber  \\
g'_{1,5}\,&=&\, 2 \binom{q-1}{2} \times 4 \; \omega_2 \,g_0^{q-3}
\,g_1 \,g_4,\\
\mbox{}  \nonumber  \\
g'_{1,6}\,&=&\, (q-1) \times 2 \, \omega_2 \; g_0^{q-2} \, g_5,
\end{eqnarray}
\end{subequations}
The contributions to $g'_2$ identical to the ones in the RF model,
shown in equations \ref{rrg2rf}. For $g_3$ we find:
\begin{subequations}
\begin{eqnarray}
g'_{3,1}\,&=&\, \omega_1\, g_0^{q-1},\\
\mbox{}  \nonumber  \\
g'_{3,2}\,&=&\, \binom{q-1}{2} \times 2 \; \omega_2 \,g_0^{q-3}
\,g_1^2 \, \\
\mbox{}  \nonumber  \\
g'_{3,3}\,&=&\, (q-1) \times 2 \; \omega_2 \,g_0^{q-2} \, g_2.
\end{eqnarray}
\end{subequations}
The contributions to $g'_4$ are:
\begin{subequations}
\begin{eqnarray}
g'_{4,1}\,&=&\, (q-1) \times 1 \; \omega_2\,g_0^{q-2}\,g_3,\\
\mbox{}  \nonumber  \\
g'_{4,2}\,&=&\, (q-1) \times 2 \; \omega_2\,g_0^{q-2}\,g_4.
\end{eqnarray}
\end{subequations}
Finally, a single contribution to $g'_5$ was found:
\begin{equation}
g'_{5,1}\,=\,(q-1) \times 2 \, \omega_2 \, g_0^{q-2}\, g_1.
\end{equation}

In the recursion relations for the RA model, we notice that the
partial partition functions $g_3$ and $g_4$ always appear in the
combination $g_3+2\,g_4$. We therefore define the following ratios for
this model: $R_1=g_1/g_0$, $R_2=g_2/g_0$, $R_3=(g_3+2\,g_4)/g_0$, and
$R_4=g_5/g_0$. The recursion relations for the ratios are:
\begin{subequations}
\begin{eqnarray}
R'_1 \,&=&\, \left[  (q-1) \, \omega_1 \,R_1+\,
 \, 6 \binom{q-1}{3} \,\omega_2 \, R_1^3 \,+\,
\right. \nonumber \\ 
&& \left. 12 \binom{q-1}{2}\,  \omega_2 \,R_1\,R_2\,+
 4 \binom{q-1}{2}\,  \omega_2 \,R_1\,R_3\, +
 \right.\nonumber \\
&&  2(q-1)\,\omega_2\, R_4 \left]\,\frac{1}{D}  \right.,\\ 
\mbox{}  \nonumber  \\
R'_2 \,&=& \,2 \,\left[\binom{q-1}{2} \,R_1^2 \,+\,
 (q-1) \, R_2 \,\right]\,\frac{\omega_2}{D},\\
\mbox{}  \nonumber  \\
R'_3\,&=&\, \left[\omega_1+2\binom{q-1}{2}\,\omega_2\,R_1^2+
  \right.\nonumber \\ 
&&   2(q-1)\,\omega_2\,(R_2+R_3) \left]\frac{1}{D} \right.,\\
\mbox{}  \nonumber  \\
R'_4\,&=&\,\frac{2\,(q-1)\,\omega_2\,R_1}{D},
\end{eqnarray}
\label{rrra}
\end{subequations}
where
\begin{eqnarray}
D\,&=&\,1\,+\, \binom{q-1}{2} \omega_1  \,R_1^2 \,+\,
 \, 6 \binom{q-1}{4}\,\omega_2 \,R_1^4 \, +\nonumber \\
\mbox{}  \nonumber  \\
&& 18 \binom{q-1}{3}\,\omega_2  \,R_1^2 \,R_2\,+\,
 6 \binom{q-1}{2}\,\omega_2  \,R_2^2\,+ \nonumber \\
\mbox{}  \nonumber  \\
&& (q-1)\,\omega_1\,R_2 \, +\, 6\binom{q-1}{3}\,\omega_2\,R_1^2\,R_3+ 
\nonumber \\
\mbox{}  \nonumber  \\
&&4\binom{q-1}{2}\,\omega_2\,R_2\,R_3\,+ 
4\binom{q-1}{2}\,\omega_2\,R_1\,R_4.
\end{eqnarray}

The partition function of the model on the Cayley tree may be obtained
if we consider the operation of attaching $q$ subtrees to the central
site of the lattice. The result for the RF model will be:
\begin{eqnarray}
Y^{(RF)}\,&=&\, g_0^q+\binom{q}{2}\,\omega_1\,g_0^{q-2}\,g_1^2\,+\,
6\binom{q}{4}\,\omega_2\,g_0^{q-4}\,g_1^4\,+ \nonumber \\
&&12 \binom{q}{3}\,\omega_2\,g_0^{q-3}\,g_1^2\,g_2\,+
4 \binom{q}{2}\,\omega_2\,g_0^{q-2}\,g_2^2.
\end{eqnarray}
For the RA model, a similar calculation leads to the partition function:
\begin{eqnarray}
Y^{(RA)}\,&=&\, g_0^q+\binom{q}{2}\,\omega_1\,g_0^{q-2}\,g_1^2\,+\,
6\binom{q}{4}\,\omega_2\,g_0^{q-4}\,g_1^4\,+ \nonumber \\
&&18 \binom{q}{3}\,\omega_2\,g_0^{q-3}\,g_1^2\,g_2\,+
6 \binom{q}{2}\,\omega_2\,g_0^{q-2}\,g_2^2\,+ \nonumber \\
&&q\,\omega_1\,g_0^{q-1}g_2\,
+\,6\binom{q}{3}\,\omega_2\,g_0^{q-3}\,g_1^2\,(g_3\,+\,2g_4)\,+
\nonumber \\
&&4\binom{q}{2}\,\omega_2\,g_0^{q-2}\,g_2\,(g_3\,+\,2g_4)\,+
\nonumber \\
&&4\binom{q}{2}\,\omega_2\, g_0^{q-2}\,g_1\,g_5.
\end{eqnarray}
In the thermodynamic limit, the solution of a model on the Cayley tree
usually shows a behavior which is quite different from the one expected
on regular lattices, since the number of surface sites represents a
non-zero fraction of the total number of sites, even in the
thermodynamic limit. We therefore study
mean values calculated at the central site of the tree. The behavior
of a model in the thermodynamic limit and in the central region of the
Cayley tree has been named the Bethe lattice solution of this model
\cite{b82}. Using the partition functions above, we then proceed
calculating the densities at the central site of the tree. The density
of monomers is given by:
\begin{equation}
\rho=\frac{P\,+\,2Q}{1\,+\,P\,+\,Q},
\label{dm}
\end{equation}
where for the RF model we have:
\begin{subequations}
\begin{eqnarray}
P^{(RF)}\,&=&\,\omega_1 \, \binom{q}{2}R_1^2, \\
\mbox{}  \nonumber  \\
Q^{(RF)}\,&=&\,\omega_2\,\left[ 6\binom{q}{4}\,\omega_2\,R_1^4\,+
\,12 \binom{q}{3}\,R_1^2\,R_2\,+ \right. \nonumber \\
\mbox{}  \nonumber  \\
&&\left. 4 \binom{q}{2}\,R_2^2 \right].
\end{eqnarray}
\label{pqrf}
\end{subequations}
For the RA model, we have the results:
\begin{subequations}
\begin{eqnarray}
P^{(RA)}\,&=&\,\omega_1 \left[ \binom{q}{2}R_1^2+q\,R_2\right], \\
\mbox{}  \nonumber  \\
Q^{(RA)}\,&=&\,\omega_2\left[6\binom{q}{4}\,R_1^4\,+\, 18\binom{q}{3}
R_1^2\,R_2\,+ \right. \nonumber \\
\mbox{}  \nonumber  \\
&& \left. 6 \binom{q}{2}\,R_2^2\,+\,
6\binom{q}{3}\,R_1^2\,R_3\,+\right. \nonumber \\
\mbox{}  \nonumber  \\
&& \left. 4\binom{q}{2}\,R_2\,R_3\,+\,
4\binom{q}{2}\,R_1\,R_4 \right].
\end{eqnarray}
\label{pqra}
\end{subequations}
It is supposed in the expressions above for the density of monomers at
the central site that the ratios $R_i$ have their fixed point values,
so that the thermodynamic limit results are obtained. Since $P$ and $Q$
are non-negative, the density of monomers will be in the interval
$[0,2]$, as expected. The probability that the central site is
occupied by a single monomer will be $\rho_1=P/(1+P+Q)$, and the
probability to find two monomers at the central site is
$\rho_2=Q/(1+P+Q)$. 

\section{Thermodynamic properties of the model}
\label{tpm}
As discussed above, the thermodynamic properties of the model on the
Bethe lattice are determined by the fixed point values of the ratios,
defined by the recursion relations, for fixed values of the parameters
$\omega_1$ and $\omega_2$. For the RF model, an inspection of the
recursion relations \ref{rrrf} shows that a fixed point with
$R_1=R_2=0$ may exist. Equations \ref{dm} and \ref{pqrf} show that in
such a phase the density of monomers vanishes, so that we may
recognize it as the non-polymerized (NP) phase. The next step is to
identify the region of the parameter space $(\omega_1,\omega_2)$ where
this phase is stable. We thus calculate the jacobian in the NP phase,
whose elements are:
\begin{equation}
J_{i,j}=\left.\frac{\partial R'_i}{\partial R_j}\right|_{R_1=R_2=0},
\label{jac}
\end{equation}
and the NP phase will be stable in the region of the parameter phase
where the absolute values of both eigenvalues of the jacobian are
smaller than 1. The jacobian matrix is diagonal and the result of this
calculation shows that the NP phase is stable when the conditions
$\omega_1<1/(q-1)$ and $\omega_2<1/[2(q-1)]$ are both satisfied. 

Another fixed point of the recursion relations for the RF model is
such that $R_1=0$ and $R_2 \neq 0$. In this phase $\rho_1$ vanishes
but $\rho_2$ not, so it is a polymerized phase in which only double
occupied (DO) sites are present. The fixed point value for $R_2$ is
given by:
\begin{equation}
R_2^{DO}\,=\, \sqrt{\frac{2 (q-1)\,\omega_2\,-\,1}{2 (q-1) (q-2)
    \,\omega_2}}, 
\end{equation}
and from \ref{pqrf} we have $P=0$ and 
\[
Q\,=\,4\,[2\,(q-1)\,\omega_2-1],
\]
which leads to the density of doubly occupied sites:
\begin{equation}
\rho_2^{DO}\,=\,\frac{8\,(q-1)\,\omega_2\,-\,4}{8\,
  (q-1)\,\omega_2\,-\,3}.
\end{equation}
The jacobian matrix is also diagonal in the DO phase, the non-zero
elements are :
\begin{subequations}
\begin{eqnarray}
J_{1,1}\,&=&\,\frac{\omega_1}{2 \omega_2} \,+\,\sqrt{\frac{2 \,(q-2)
    \,\left[2 (q-1)\,\omega_2\,-\, 1\right] \, }{(q-1)\,\omega_2}},\\
J_{2,2}\,&=&\,\frac{1}{(q-1)\,\omega_2}\,-1.
\end{eqnarray}
\end{subequations}
The DO phase is stable if the absolute value of both eigenvalues of
the jacobian is less or equal to one, which leads to the conditions:
\begin{equation}
\frac{1}{2 (q-1)} \,\leq \omega_2 \, \leq \,\frac{2 (q-2)}{(q-1) (4
  q-9)},
\label{sldo1}
\end{equation}
and for a value of $\omega_2$ in the range above, we have:
\begin{equation}
\omega_1(\omega_2)\,\leq\,2 \omega_2\,-\,
2 \sqrt{\frac{2\,\left[2 (q-1)\,\omega_2\,-\,1\right]
\,(q-2) \,\omega_2}{(q-1)}}.
\label{sldo2}
\end{equation}
It may be observed that this phase has a non-vanishing region of
stability for any value of the lattice coordination number. For large
values of $q$ we may expand the limits of the stability interval in
powers of $1/q$, finding that:
\begin{eqnarray*}
 \frac{1}{2 (q-1)}&\sim_{q\rightarrow \infty }& \frac{1}{2 q} \left(1 + 
 \frac{1}{q}+\ldots\right), \\
\mbox{}  \nonumber  \\
  \frac{2 (q-2)}{(q-1) (4 q-9)}&\sim_{q\rightarrow \infty }& \frac{1}{2 q} 
  \left(1 + \frac{5}{4 q}+\ldots\right) .
\end{eqnarray*}
We conclude that the NP-DO phase transition
is continuous, at $\omega_2=1/[2(q-1)]$. For $\omega_2>1/[2(q-1)]$ the
other stability limit of the DO phase is given by expression
\ref{sldo2}. 

A third fixed point of the recursion relations \ref{rrrf}, which we
may associate to the regular polymerized phase (P), corresponds to
both ratios at non-zero values. The analytic study of this fixed point
is harder, but with the help of a formal algebra software we may
conclude that the stability limit of the DO phase is never
coincident with the stability limit of the P phase, and thus the DO-P
transition is always of first order. The stability limits of the NP
and the P phase are the same for $\omega_2<1/[2(q-1)^2]$, henceforth
the transition between these phases is continuous in this region. For
larger values of $\omega_2$, this transition becomes discontinuous. A
tricritical point is found at:
\begin{subequations}
\begin{eqnarray}
\omega_1^{TC}\,&=&\,\frac{1}{q-1},\\
\omega_2^{TC}\,&=&\,\frac{1}{2(q-1)^2}.
\end{eqnarray}
\label{tcp}
\end{subequations}

\begin{figure}
\begin{center}
\includegraphics[height=6.0cm]{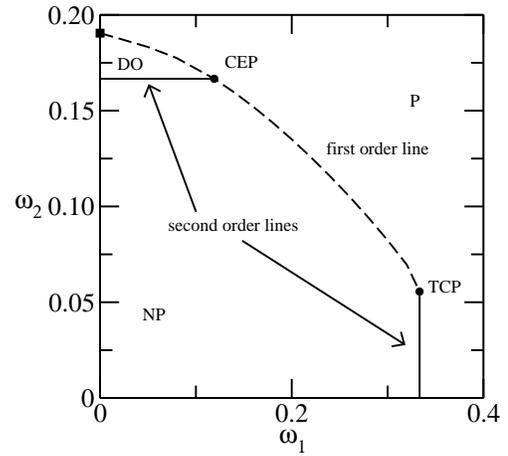}
\caption{Phase diagram for the RF model on a Bethe lattice with
  $q=4$. Full lines are continuous transitions and on the broken lines
  two phases coexist. NP denotes the non-polymerized phase
  ($\rho_1=\rho_2=0$), DO the double occupancy phase ($\rho_1=0,\rho_2
  \neq 0$, and P the regular polymerized phase ($\rho_1,\rho_2 \neq 0$).}
\label{f3}
\end{center}
\end{figure}

The first-order transition lines may be found through a Maxwell
construction \cite{sms96}, but this procedure may be difficult to
converge numerically if more than two phases are present. For
simplicity, we thus adopted an alternative procedure \cite{p02},
iterating the recursion relations starting with the physical values
for the partial partition functions for a subtree of generation
``zero'', which are $g_0^{(0)}=1$, $g_1^{(0)}=\omega_1$, and
$g_2^{(0)}=\omega_2$. With these starting values, the recursion
relations do converge to the stable thermodynamic phase and therefore
we may obtain the first-order transition lines. In Fig. \ref{f3} we
show the phase diagram of the RF model for $q=4$. The continuous lines
are second order transitions and at the broken lines two distinct
phases coexist. The continuous NP-P transition lines ends at a
tricritical point located at $\omega_1=1/(q-1)=1/3$ and
$\omega_2=1/2(q-1)^2=1/18$. The 
second order NP-DO transition ends at a critical endpoint at
$\omega_1=0.1190365$ and $\omega_2=1/6$ ($\omega_1$ was determined
numerically). It is worth mentioning that there is no discontinuity at
the point where the P-DO coexistence line touches the $\omega_1=0$
axis, denoted by a square in the phase diagram. Also, the fact that
the DO phase becomes unstable as $\omega_1$ is increased for fixed
$\omega_2$ may be understood noting that although it is energetically
more favorable to have doubly occupied sites at high values of
$\omega_2$, a non-zero density of simply occupied sites increases the
entropy. 

\begin{figure}
\begin{center}
\includegraphics[height=6.0cm]{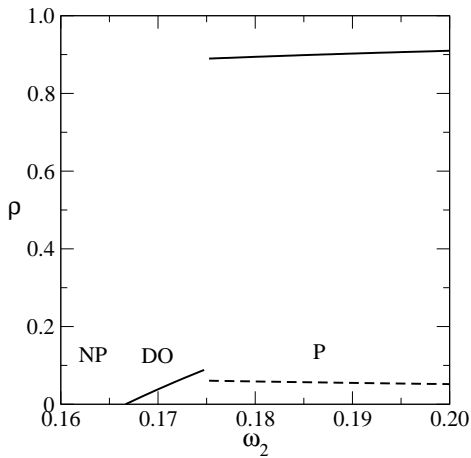}
\caption{Densities for the RF model with $q=4$ as functions of
  $\omega_2$, for $\omega_1=0.09$. The full line is the density of
  double occupied sites ($\rho_2$) and the dashed line corresponds to
  the density of single occupied sites ($\rho_1$).}
\label{f4}
\end{center}
\end{figure}

In Fig. \ref{f4} the densities of single- and double occupied sites
are shown as functions of $\omega_2$ for a fixed value
$\omega_1=0.09$, again for a lattice with $q=4$. 
For low values of $\omega_2$ the NP phase is stable. At
$\omega_2=1/2(q-1)=1/6$, a continuous transition to the DO phase
occurs, followed by a discontinuous transition to the P phase at
$\omega_2\,\approx\,0.17503$.  

Turning our attention to the more general RA model, we notice from the
recursion relations \ref{rrra} that the
non-polymerized phase has now the fixed-point values $R_1=R_2=R_4=0$
and
\begin{equation}
R_3=\frac{\omega_1}{1-2(q-1)\omega_2}.
\end{equation}
These values for the ratios actually correspond to $\rho=0$, as may
be seen using equation \ref{dm} for the density of monomers. The
stability limit of the NP phase is obtained requiring that the largest
eigenvalue of the jacobian, calculated at the fixed point values of
the ratios above, has an absolute value equal to 1. This leads to the
following condition:
\begin{equation}
\omega_1^{(NP)}=\frac{[1-2(q-1)\omega_2]^2[1+2(q-1)\omega_2]}
      {1-2\omega_2}.
\end{equation}

We may search for a DO fixed point, which should have the values
$R_1=R_4=0$ and $R_2,R_3 \neq 0$. However, writing the equation for
the fixed point value of the ratio $R_2/R_3$, we reach the unphysical
conclusion $R_2=-\omega_1/[2(q-1)\omega_2]$, indicating that such a
phase does not appear in the $RA$ model for $\omega_1 \neq 0$, since
ratios of partial partition functions should be non-negative. We may
have some physical understanding of the non-existence of a region
where the DO phase is stable in the $RA$ model 
if we notice that the DO phase found in the $RF$ model
is composed by linear chains with double polymer bonds between first
neighbor sites occupied by two monomers, which are constrained to have
endpoints at the surface of the tree. We may notice that in the $RF$
model, for $\omega_1=0$, we have a continuous transition from the NP
phase to the DO phase at $\omega_2=1/[2(q-1)]$. If we take into
account that at each lattice site there are two ways to connect the
bonds to the monomers on this site for a linear DO polymer, this
corresponds exactly to the well known result that the polymerization
transition for linear chains on a Bethe lattice occurs when the
statistical weight of a 
monomer is equal to $1/(q-1)$ \cite{sw87}, as may be also seen in our
results for the $RF$ model in the particular case $\omega_2=0$. If we
allow immediate 
self-reversals, these chains may end at any step inside the
tree. Therefore, this corresponds to allow for {\em endpoints} in the
problem of polymerization of linear chains, and it is known that
this effect eliminates the polymerization transition. It is possible
to map the equilibrium polymerization model into the $n$-vector model
of magnetism in the formal limit $n \to 0$, and if this is done, the
statistical weight of monomer at the end of chains is proportional to
the magnetic field in the $n$-vector model \cite{wp81a}. The continuous
ferromagnetic 
transition in the $n$-vector model exists only when no magnetic field
is applied. Accordingly, no polymerization transition is found for
nonzero statistical weight of monomers at endpoints of chains.

\begin{figure}
\begin{center}
\includegraphics[height=6.0cm]{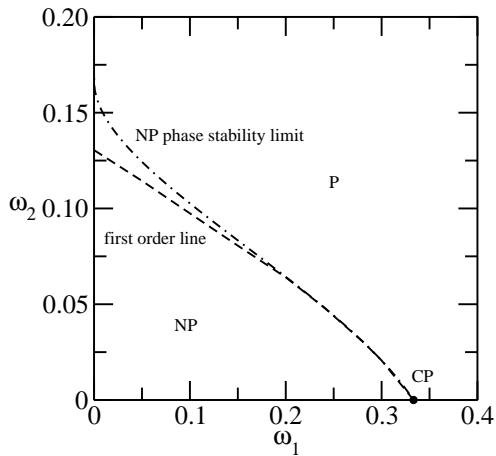}
\caption{Phase diagram for the RA model on a Bethe lattice with
  $q=4$. Only the non-polymerized (NP and the regular polymerized (P)
  phases appear. For non-zero $\omega_2$, the phase transition is always
  discontinuous. The stability limit of the NP phase is shown as a
  dash-dotted line.}
\label{f5}
\end{center}
\end{figure}

The phase diagram for the RA model is shown in Fig. \ref{f5}. The
coexistence line between the NP and P phases was again found iterating
the recursion relations \ref{rrra} with initial values $g_0^{(0)}=1$,
$g_1^{(0)}=\omega_1$, $g_2^{(0)}=\omega_2$, $g_3^{(0)}=\omega_1$,
$g_4^{(0)}=0$, and $g_5^{(0)}=\omega_2$. In the RA model, the NP-P
transition was found to be continuous only at vanishing values of
$\omega_2$, so that the phase diagram is qualitatively different from
the one found for the RF model. At $(\omega_1=1/(q-1),\omega_2=0)$ we
have, therefore, a critical point and no tricritical point is present
in the phase diagram. 

\section{Conclusion}
\label{con}
A model of self- and mutually avoiding chains which allows for up to
$K=2$ monomers per lattice site was solved on the Bethe lattice with
general coordination number $q$. Endpoints of the chains are allowed
only at the surface of the tree. Following \cite{kpor06}, besides the
general model (RA), a restricted version (RF) where immediate
self-reversals of the chains are forbidden, was also
considered. Although in some respects this model resembles models for
branched polymers which were already studied on the Bethe lattice
\cite{bs95}, the multiplicities of the contributions to the partition
functions are different, since each  configuration has to be a self
avoiding walk passing through all monomers placed on the lattice. The
statistical weight of a configuration is
$\omega_1^{N_1}\,\omega_2^{N_2}$, where $N_1$ ($N_2$) is the number
of lattice sites with 1 (2) monomers. For low values of $\omega_1$ and
$\omega_2$, a non-polymerized phase is stable for both models. The
phase diagram for the RF model displays two distinct polymerized
phases, one of which (DO) with double occupied sites only. In this
model, the transition between the NP phase and the regular polymerized
(P) phase may be of first or second order, and therefore a tricritical
point is found, which could be associated with the
$\theta$-point. Also, the second-order transition between the NP and
the DO phases ends at a critical endpoint. In the solution of the more
general RA model only one polymerized phase is found, and the
transitions between it and the NP phase is always discontinuous for
nonzero $\omega_2$. Thus, no tricritical point is found in this
case. In both models no dense phase was found, similar the one
found for the ISAW model with interacting bonds on the $q=4$ Husimi
lattice \cite{sms96,p02,sscm04}.

One conclusion of this study is that both versions of the model have
qualitatively different phase diagrams on the Bethe lattice. It is
worth mentioning that simulations of the model for $K=3$ also
presented results which were strongly dependent upon the restrictions
of the model \cite{kpor06}, although no direct comparison can be done
between our results and the simulations, since they correspond to
different values of the maximum number of monomers per lattice site
$K$. Although at least for the RF model a tricritical point is found
which may be associated to the collapse transition of the chains, if
we consider the location of this point we are lead to an
inconsistency. We may adopt a physically reasonable parametrization of
the two statistical weights, stating that $\omega_1=z=\exp(\beta \mu)$
is the fugacity of one monomer and $\omega_2=z^2 \omega$, where
$\omega$ is the Boltzmann factor associated with the interaction of
the two monomers placed on the same site (interactions between
monomers at different sites are not considered in the model). The
tricritical point, whose location is given by the equations \ref{tcp},
corresponds to a fugacity $z=1/(q-1)$ and
a Boltzmann factor $\omega=1/2$. Since $\omega<1$, the collapse
transition would occur for {\em repulsive} interactions between
monomers occupying the same site. For the ISAW model on the Bethe
lattice, the tricritical point is located at $z=1/(q-1)$ and
$\omega=(q-1)/(q-2)>1$ \cite{ss90}, in the attractive region. It
should be mentioned, however, that the solution of the ISAW model
(interaction between monomers) on the Husimi tree with $q=4$ leads to a
tricritical point at $\omega \approx 1.54$ \cite{p02}, which is
somewhat higher 
than the Bethe lattice result. It may be interesting to see if the
solution of the RF model on the Husimi tree (which is supposed to
be closer to the solution on regular lattices), displays a tricritical
point in the physically expected attractive region. We are presently
working on this problem.

\section*{Acknowledgements}

This work was partially financed by project
Pronex-CNPq-FAPERJ/\-171.168-2003. PS acknowledges the hospitality of
Universidade Federal Fluminense, where part of this work was done, and
partial financial support  of the Argentinian agencies SECYTUNC and
CONICET.

\end{document}